# Imaging Mechanism for Hyperspectral Scanning Probe Microscopy via Gaussian Process Modelling

Maxim Ziatdinov,[1,2] Dohyung Kim,[3] Sabine Neumayer,[1] Rama K. Vasudevan,[1] Liam Collins,[1] Stephen Jesse,[1] Mahshid Ahmadi,[3] and Sergei V. Kalinin[1]

[1] Center for Nanophase Materials Sciences, Oak Ridge National Laboratory, Oak Ridge, TN 37831, USA

[2] Computational Sciences and Engineering Division, Oak Ridge National Laboratory, Oak Ridge, TN 37831, USA

[3] Joint Institute for Advanced Materials, Department of Materials Science and Engineering, University of Tennessee, Knoxville, TN 37996, USA

We investigate the ability to reconstruct and derive spatial structure from sparsely sampled 3D piezoresponse force microcopy data, captured using the band-excitation (BE) technique, via Gaussian Process (GP) methods. Even for weakly informative priors, GP methods allow unambiguous determination of the characteristic length scales of the imaging process both in spatial and frequency domains. We further show that BE data set tends to be oversampled, with ~30% of original data set sufficient for high-quality reconstruction, potentially enabling the faster BE imaging. Finally, we discuss how the GP can be used for automated experimentation in SPM, by combining GP regression with non-rectangular scans. The full code for GP regression applied to hyperspectral data is available at https://git.io/JePGr.

Over the last three decades, scanning probe microscopy (SPM) has emerged as a primary tool for characterization of structure and functionality at the nanometer and atomic scales. Following the initial demonstration of topographic imaging in contact mode,[1] the intermittent and non-contact topographic imaging modes, as well as SPM modes for electrical,[2] magnetic,[3] mechanical,[4] and electromechanical imaging[5,6] and spectroscopy[7] followed. This rapid development of imaging modalities implemented on these benchtop tools have opened the nanoworld for exploration, and was one of the key elements underpinning the explosive growth of nanoscience, nanotechnology, and many areas of fundamental and applied research.[8]

All SPM modes rely on the concept of the local probe interacting with the surface. Scanning the probe allows sequential measurements of materials response over a spatial grid, giving rise to images or hyperspectral images, i.e. multidimensional SPMs or SPM spectroscopic imaging modes. The data acquisition process in SPM can hence be represented as the combination of two elements – the SPM engine, or the excitation/detection scheme, and the spectroscopic mode. The spectroscopic modes define the sampling of the parameter space of interest at each spatial location, for example, classical force-distance and current-voltage curve mappings in atomic force microscopy[9] and scanning tunneling microscopy, and complex time- and voltage spectroscopies in Piezoresponse Force Microscopy and Electrochemical Strain Microscopy.[10-12] In modern SPMs, the parameter space is usually sampled sequentially, albeit this limitation is not rigid.

The SPM engine defines the response at a single point in the parameter space, i.e. the nature of response at a single voxel. Classical SPM engines typically employ a combination of sinusoidal excitation and lock-in detection, e.g. as used in classical amplitude detection SPM, or a combination of sinusoidal excitation with the phase locked loop detection in frequency detection schemes. These two engines yield (multimodal) scalar information, i.e. two response values value per pixel. Examples of vector detection engines are the band excitation,[13] exciting and detecting multiple frequencies in parallel, and harmonic intermodulation methods[14] that detect the mixing harmonics between the two excitation signals. In both cases, the engine compresses the data stream from the detector to a single parameter or a set of parameters corresponding to response vector components. Finally, more complex detection schemes such as G-Mode SPM[15-17] are based on detection and storage of the full data stream from the detector, obviating the data compression stage during scanning, after which in-depth data analysis can be performed. Both BE and intermodulation, and G-Mode detection were extended to a broad variety of SPM modes including topographic imaging, magnetic and electrostatic force imaging,[18] piezoresponse force microscopy,[19] and is expected to be universally applicable to all SPM modes.

The proliferation of the BE, intermodulation, and G-Mode SPMs and their nascent adoption by the commercial vendors necessitates understanding basic image formation mechanisms in these techniques. To date, most of such analyses were based on the physics-based models, where the known (or postulated) physics of the imaging process was used to transform the high-dimensional data to a number of reduced, ideally material-specific (i.e. independent of imaging system), parameters. The examples of such analysis include the simple harmonic oscillator fit in the BE methods,[20] or reconstruction of the intermodulations harmonics in IM.[14,21] Here, the measured resonance frequencies and force-distance curves describing salient features of tip-surface interaction can then be transformed into effective Young-moduli.

At the same time, of interest is the amount of information contained in the multidimensional data as determined from a purely information theory viewpoint, as well as approaches to compress and visualize it for exploratory data analysis. This information will provide both insight into the fundamentals of imaging mechanisms and tip-surface interaction, can be correlated with materials structure to yield insight into materials behaviors, and suggest strategies for automated experimentation. Previously, hyperspectral datasets were explored using multivariate statistical methods such as principal component analysis (PCA),[22] more complex methods such as non-negative matrix factorization that allow for certain physics-based constraints,[23] or non-linear autoencoders.[24] However, these methods are based exclusively on the analysis of the spectral dimensions, whereas spatial correlations are explicitly ignored. In other words, the endmembers of spectral unmixing do not depend on the relative positions of the spatial pixels. Correspondingly, analysis of spatial features in the loading maps was used as a way to infer the understanding of the system.[23] Alternatively, neural network based algorithms were suggested as an approach to identify the data based on labeled examples[25] or to extract the parameters of a theoretical model.[26,27]

Here, we explore the imaging mechanisms in the band excitation Piezoresponse force microscopy using the Gaussian Process regression.[28-32] This method allows to explore the data structure from purely information theory perspective simultaneously in the spatial and parameter space. We apply this Bayesian machine learning approach to determine the characteristic length scale of the phenomena, information content in the hyperspectral images, and suggest the strategies for automated experimentation based on exploiting sampling of space where maximum uncertainty is predicted.

The BE PFM measurements were performed on an Asylum Cypher microscope using an in-house built BE controller. As a model system, we use an epitaxial BiFeO$_3$ thin film of 100 nm thickness. The characteristic surface topography is shown in Figure 1 (a). The BE data can be fitted to a SHO model to extract amplitude, phase, resonance frequency and Q-factor maps as shown in Figure 1 (b-e). These images clearly indicate a ferroelectric domain structure as expected for this material. Figure 1 (f) shows the amplitude zoomed-in at the top left corner of panel (b).

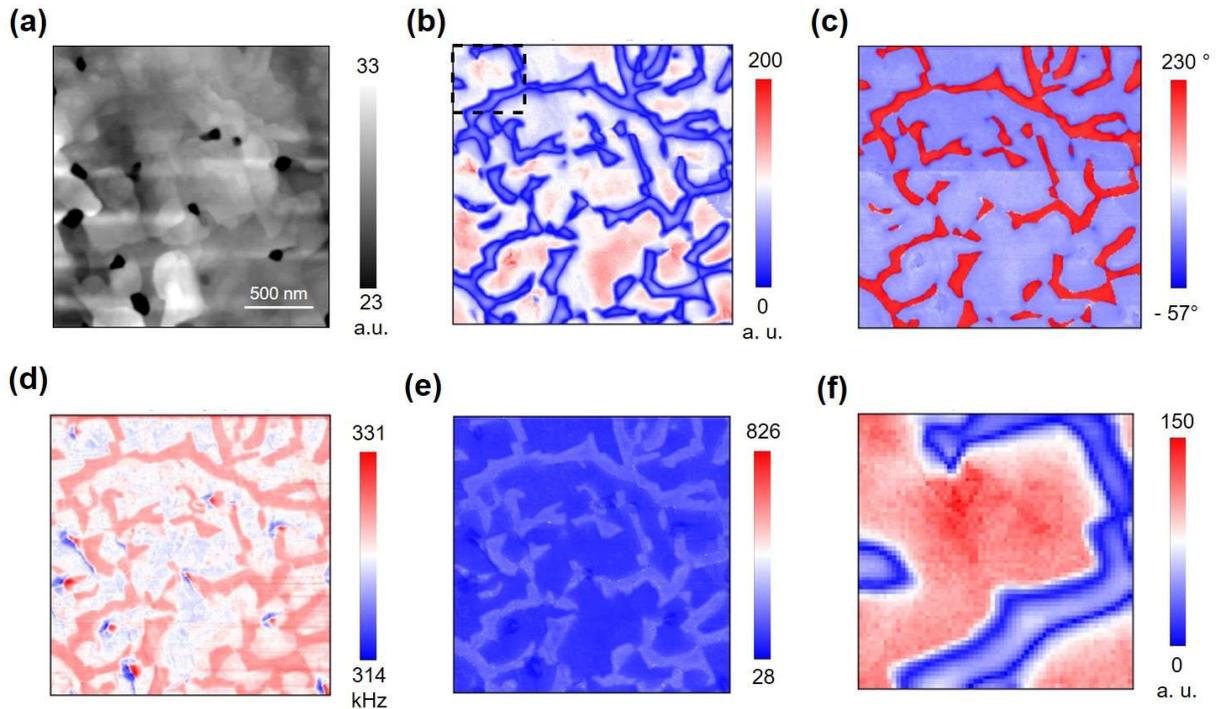

**Figure 1.** BE-PFM images in the BiFeO$_3$ epitaxial film. (a) Topographic map. (b) BE-PFM amplitude, (c) phase, (d) resonance frequency, and (e) Q-factor maps. (f) Zoomed in amplitude image at top left corner.

To explore the information distribution in the BE data set, we employ the Gaussian Process (GP) method.[31,32] GP generally refers to an indexed collection of random variables, any finite number of which have a joint Gaussian distribution. More simply, a GP is a distribution over functions on a given domain and can be used for approximating continuous nonlinear functions.

GP can be completely specified in terms of its mean function and covariance function (also referred to as kernel function). A common application of GP in machine learning is a GP regression analysis where one estimates an unknown function given noisy observations $\mathbf{y} = (y(\mathbf{x}_1), \ldots, y(\mathbf{x}_n))^T$ of the function at a finite number of points $X = \{\mathbf{x}_1, \ldots, \mathbf{x}_n\}$. The marginal likelihood of the data is conditioned on the kernel hyperparameters $\boldsymbol{\theta}$ as following:

$$\log p(\mathbf{y}|X, \boldsymbol{\theta}) \propto -\frac{1}{2}\log|\mathbf{K}_{\boldsymbol{\theta}} + \sigma^2 \mathbf{I}| - \frac{1}{2}\mathbf{y}^T(\mathbf{K}_{\boldsymbol{\theta}} + \sigma^2 \mathbf{I})^{-1}\mathbf{y} \tag{1}$$

where the first and second terms can be interpreted as the hyperparameters learning and the data-fit, respectively.[31] Here $\sigma^2$ is noise variance. Because the application of GP models to large datasets is intractable, we adapt a sparse GP regression method for constructing an approximation using a subset of observations called inducing points.[33] The inducing points are optimized together with kernel hyperparameters during model training. Maximizing the number of inducing points generally yields more accurate results, albeit at the cost of computation time and memory.

It is important to know that the signature aspect of GP method is that observations at different locations are assumed to be linked via the kernel function, defining the connection between the dissimilar locations. The kernel function can be either defined *a priori*, i.e. from the known physics of the system or additional information, or can be treated as a hyperparameter. In the latter case, the functional form of the kernel is defined and the corresponding parameters are determined as a part of the fitting process. Here, we note that the kernel parameters determined self-consistently as a part of the regression process should provide robust information on the image formation mechanism in the technique, and explore this proposition below.

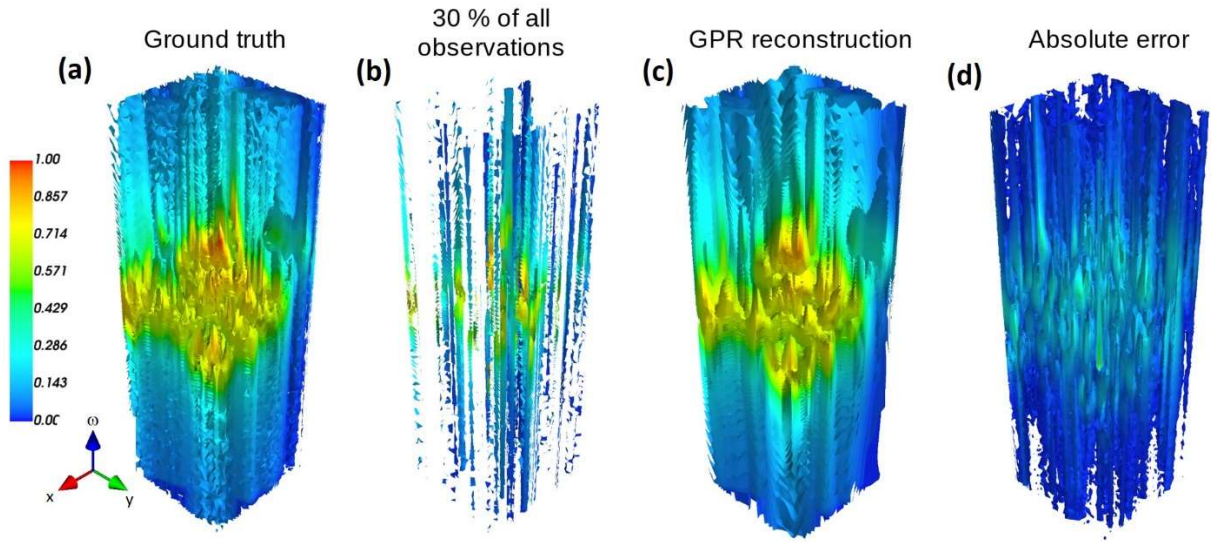

**Figure 2.** Illustration of GP application to 3D spectroscopic dataset (the dataset dimensions are 32 x 32 x 102). Original image (ground truth) in (a) is corrupted by removing 70% of observations (measured spectroscopic curves) as shown in (b). The GP regression (GPR) is then used to reconstruct the signal (c). The absolute error is shown in (d). Note that the absolute error may be misleading when the ratio of signals of interest doesn't change (see Fig. 3).

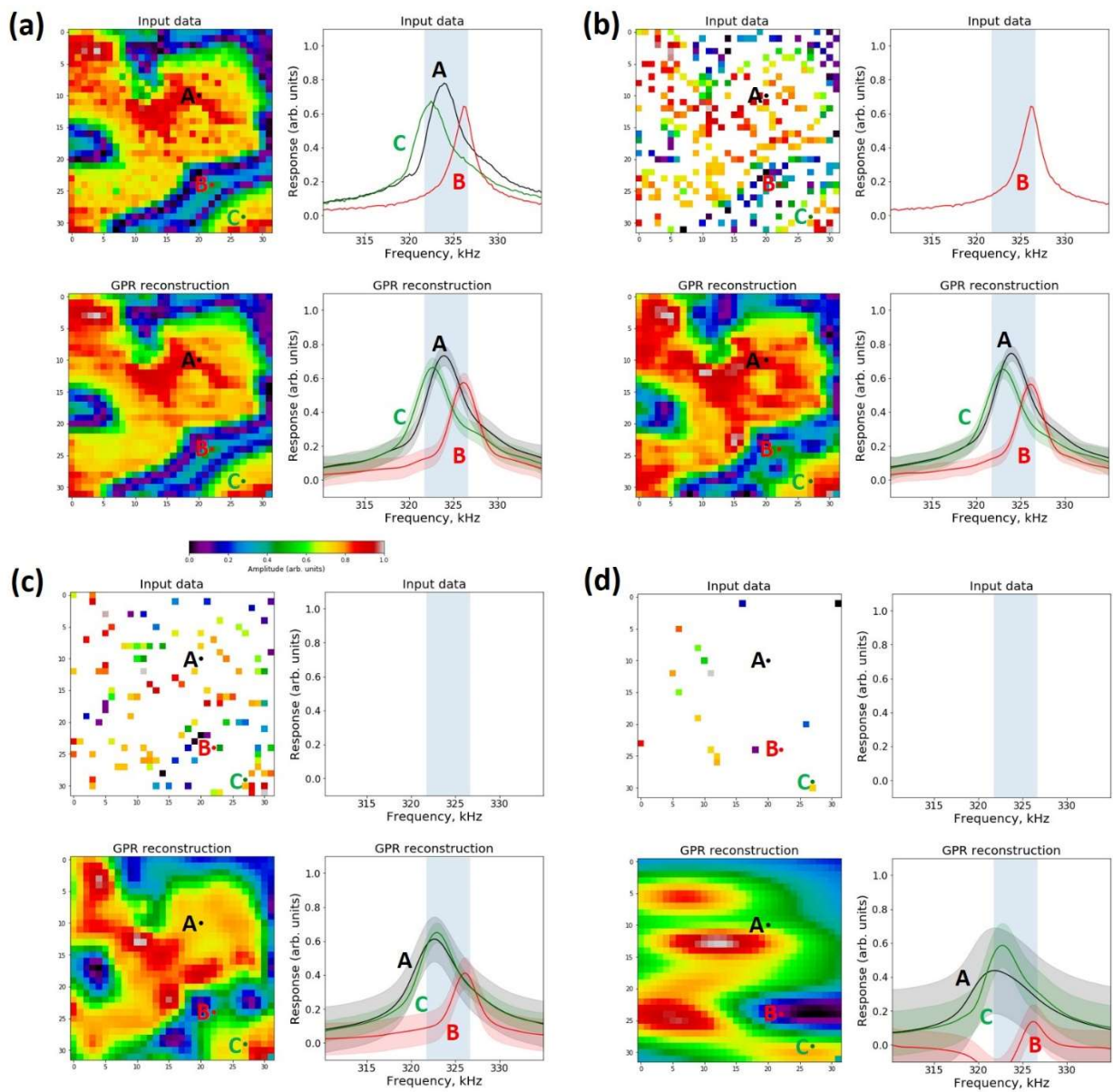

**Figure 3.** Selected 2D representations corresponding to the application of GP regression to original (full) data (a) and data corrupted by removing 70% (b), 90% (c) and 99% (d) of observations. In each panel, the top row shows the input data (2D slices of hyperspectral data average over selected frequency range and spectroscopic curves extracted at A, B and C points) and the bottom row shows the model output (the same averaged slices and the reconstructed curves from the same locations as in the input data). The vertical blue span in the plots indicate a slice used for 2D image plots (the same in all four panels). The spatial resolution in all the images is ~9.76 nm/px. The full 3D datasets are available from the notebook.[34]

To explore the applicability of the GP processing in BE, we first demonstrate its potential for reconstruction from *partial* data and subsequently show how exploratory analysis using information about maximum uncertainty as a guide for selecting the next measurement point can be performed. We selected a 32 x 32 x 102 subset of the original hyperspectral data (see Figure 1(f)) in order to reduce computational time/costs and to make it easier to reproduce our results without a need to use high-performance computing. Our GP analysis code tailored towards the analysis of 2D and 3D image and spectroscopic data is based on Pyro probabilistic programming language.[32] We used a fixed number (1500) of inducing points (determined based on the limits of GPU memory) for reconstruction of individual data sets, while for the "sample exploration" problem the number of inducing points was set to 5 % of the overall data points. We also provide an executable Jupyter notebook for reproducing the paper's results.[34] The notebook can be executed either using a standard Google Cloud Platform virtual machine with NIVIDIA's Tesla P100 GPU and 15 GB of RAM (running the notebook one time from top to bottom costs ~2 USD) or in Google Colab with NVIDIA's Tesla K80 GPU, which is free of charge but may require significantly longer computational times.

Here, the original BE data set is considered to be the "known" ground truth. A part of the data is removed, creating the artificial data set. The GP regression is used to reconstruct the full data set, and the reconstruction error is evaluated. This process is illustrated in Figure 2, showing the original data set as a response in the *x*, *y*, and frequency space (Figure 2 (a)), the reduced data set (Figure 2 (b)), the reconstructed data set (Figure 2 (c)), and the absolute error (Figure 2 (d)). Here, the 'Matern52' kernel is used.[31] The reconstruction error shown in Figure 2 (d) generally does not exceed ~20% despite the fact that 70% of the data was eliminated.

To explore the robustness of this approach, we explore the veracity of reconstruction for the various degrees of image reduction. The 2D representation of the reconstruction of the BE data set for the full data set and for the data sets with removal of 70%, 90%, and 99% of the original data is shown in Figure 3. Even for the data reduction by 90%, the reconstructed data sets maintain the characteristic features of the response, including both the general domain configuration and the behavior of the amplitude-frequency curves. Note that GP process yields not only estimated response values, but also the confidence intervals at the given point in the parameter space, thus allowing for the formulation of optimal strategies for experiment automation, as will be explored later. Overall, we conclude that the BE data is strongly oversampled, and even without strong physics-specific priors the GP process allows reconstruction of data from partial observations.

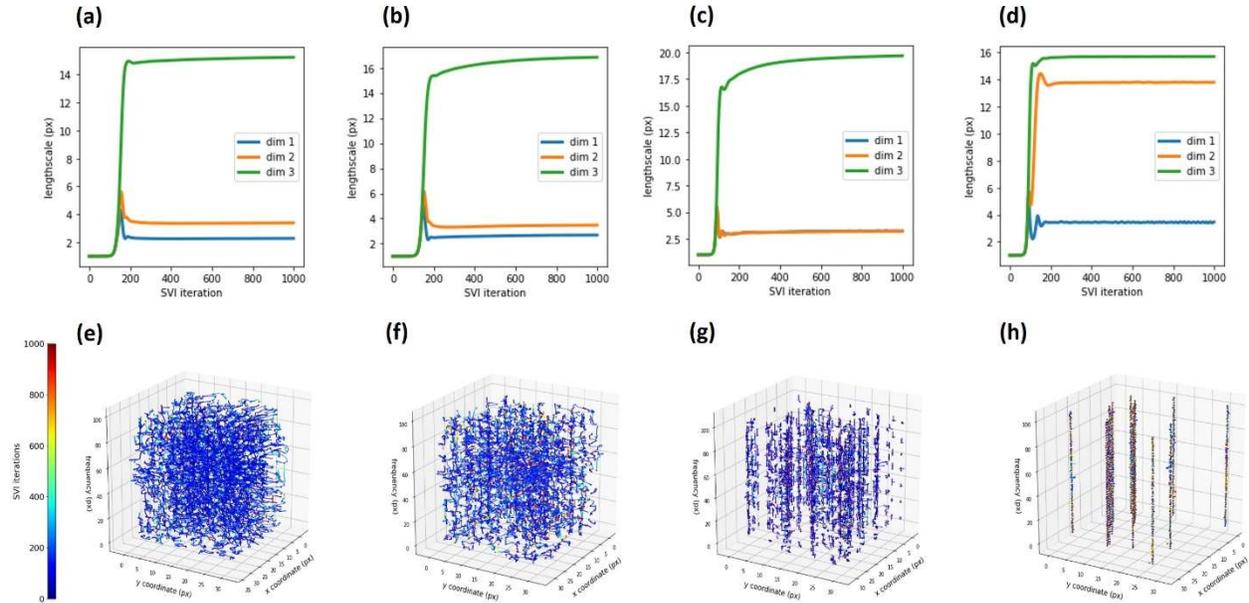

**Figure 4.** Evolution of kernel lengthscales (a-d) and inducing points (e-h) during the stochastic variational inference (SVI) based model training for full data set (a, e), and data corrupted by removing 70 % of observations (b, f), 90 % of observations (c, g) and 99 % of observations (d, h). (a) The first two dimensions in (a-d) (dim 1 and dim 2) correspond to *x* and *y* coordinates, whereas the third dimension (dim 3) corresponds to frequency. The kernel lengthscales define the spatial resolution of the technique (assuming atomically thin domain wall width) in the spatial domain, and the width of the resonance peak in the frequency domain.

We further note that the key element of GP process is that it yields the insight into the structure of the data via the kernel parameters. Here, we use the weakly informative 3D Gaussian kernel, with the characteristic length scales determined as a part of regression process. These length scales hence define the characteristic resolution in the spatial and frequency domains, and do so in robust (with respect to noise) fashion. These behaviors are illustrated in Figure 4 (a-d) for the full data set and for the partially reduced data sets discussed in Figure 3. Note that for all cases discussed in Figure 3, the frequency length scale converges to the similar value given by the width of the BE peak. For spatial length scales, the analysis of the full data set allows to establish characteristic spatial resolution as the kernel length scale. This length scale in turn provides robust estimate of the characteristic length scale of tip-surface interactions. We also showed trajectories of inducing points for each case, which were selected as a subset of the input data points by taking every *n*-th point from the original dataset. Here, *n* was adjusted such that the total number of inducing points remained the same (1500, based on GPU memory limits) for all 4 scenarios. Notice that most inducing points in Figure 4 (e - g) stop their motion after ~500 stochastic variational inference (SVI) iterations and that for data set with 99 % of observations removed all the inducing points remain mostly at their original location. While the inducing points do not have a physical meaning, understanding their evolution during the optimization process is important for applying GP models to larger hyperspectral datasets and to the experiment automation.

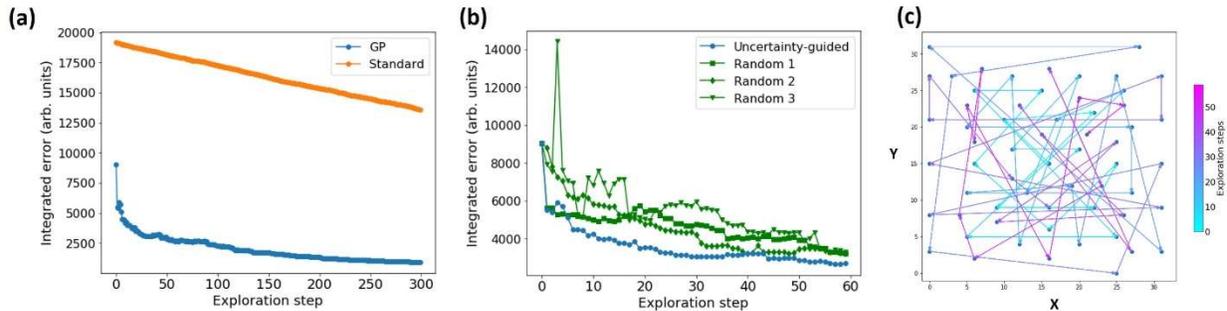

**Figure 5.** Simulation of the GP-guided experimental measurements. (a) The data is real (experimental). (a) The integrated absolute error versus exploration steps for measurements with ('GP') and without ('standard') GP regression-based reconstruction. (b) Comparison of the integrated absolute error when the next measurement point is selected using the maximal uncertainty in GP reconstruction *versus* when the next measurement point is selected randomly (for 3 different random seeds). (c) Exploration path in *xy* coordinates for the first 60 steps.

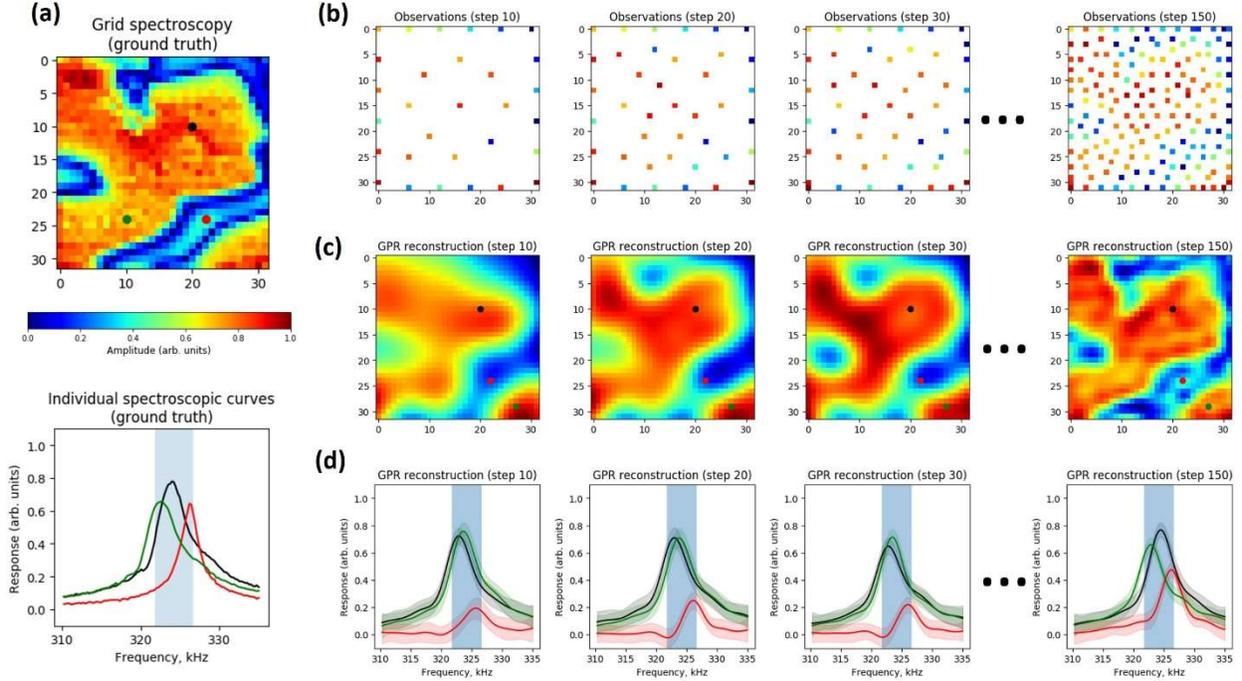

**Figure 6.** Simulation of the GP-guided experimental measurements. The data is real (experimental (a) Ground truth data in 2D representation for the selected frequency range and the individual spectroscopic curves from the selected locations. (b) Input data updated with a new single measurement at each step. (c) GP regression-based reconstruction of the entire field of view, integrated over a frequency range shown by a vertical blue span in (3). (d) Reconstructed individual spectroscopic curves from the locations shown by black, red and green dots in (c).

We finally demonstrate that GP regression can be used for guiding the actual measurements. Here, we start with just a few measured points along each edge in the *xy* plane (~1% of all the observations). A single exploration step consists of i) performing a GP regression, ii) doing a single "measurement" in a point with maximum uncertainty and iii) updating the previous input data with the data points associated with this measurement. Here we demonstrate this approach using a "synthetic" experiment, that is, we use the data from the actual but already completed experiment, with 99% of all the observations removed, which allows us estimating the absolute error at each step. As one can observe from Figure 5a, the absolute error is rapidly decreasing during the first ~30 exploration steps. The error is much higher when the next measurement point is selected just randomly, as demonstrated in Figure 5b. We noticed that if the edge points are not "opened" in the beginning, the algorithm typically spends the first ~10 iterations (for this particular data dimensions) on measuring the edges since this is where the kernel diverges (and not because this is related to any sample properties) and then moves to the regions deeper inside the field of view. The selected inputs and outputs for this exploration range are shown in Figure 6. Interestingly, one can get a good understanding of the sample domain structure already after 20-30 steps of such an autonomous experiment, which suggests that this approach can significantly reduce the data acquisition time. The remaining steps are typically spent on "refining" the uncovered structures (compare, for example, the third and the last column in Figure 6b-d). The current approach can be improved by introducing a "cost function" determining which objects are of real physical interest (that, is "worth exploring"), in addition to a pure uncertainty-based exploration.

To summarize, we have explored the applicability of the Gaussian Process regression with the weakly-informative priors for the analysis of the band excitation data on example of Piezoresponse Force Microscopy. Here, we explore the signal in the 3D (x, y, frequency) parameter space. Even for the weakly informative priors, the GP methods allows to unambiguously determine the characteristic length scales of the imaging process both in spatial and frequency domain. We further show that BE data set tend to be oversampled, with ~30% of original data set sufficient for high-quality reconstruction.

We further note that this analysis points at strong potential of GP for the development of the automated experiments, where the measurement points are chosen based on the results of previous measurements.[35-37] Here, the Bayesian uncertainty along with the target-driven criteria can be used for balancing of the exploratory and exploitation activity. Furthermore, we believe that the analysis can be strongly improved with the addition of physics-based priors to reconstruction.

Acknowledgements:
The authors would like to thank Amit Kumar (Queen's University Belfast) and Dipanjan Mazumdar (Southern Illinois University) for providing the $BiFeO_3$ sample. Research was conducted at the Center for Nanophase Materials Sciences, which is a DOE Office of Science User Facility (MZ, RKV, LC, SJ, SVK). Part of the BE SHO data processing and experimental setup were supported by the U.S. Department of Energy, Office of Science, Basic Energy Sciences, Materials Science and Engineering Division (SMN). M.A. and D. K. acknowledge support from CNMS user facility, project # CNMS2019-272.

Competing interests:

The authors declare no Competing Financial or Non-Financial Interests.

Data Availability:

The full code and data are available at https://git.io/JePGr.

Author contributions:

SVK proposed the concept and led the paper writing. MZ wrote the code for GP-based analysis of hyperspectral SPM data, performed the analysis and co-wrote the paper. DK collected the experimental data, with assistance from SN and LC. SJ and LC configured the original experimental set-up. RVK, MA and SJ assisted with data interpretation and paper writing.